\documentclass[useAMS]{mn2e}
\usepackage{epsfig,graphics}
\usepackage{amsmath}

\title[Intra-night variability of CTA~102]{Intra-night variability of the blazar CTA~102 during its 2012 and 2016 giant outbursts}

\author[Bachev et al.]{R. Bachev$^{1}$\thanks{E-mail: bachevr@astro.bas.bg}, 
V. Popov$^{2}$, 
A. Strigachev$^{1}$,
E. Semkov$^{1}$,
S. Ibryamov$^{1, 3}$, 
B. Spassov$^{1}$,
\newauthor G. Latev$^{1}$, 
R. V. Mu\~{n}oz Dimitrova$^{1}$, 
S. Boeva$^{1}$
\\
\\
$^{1}$ Institute of Astronomy, Bulgarian Academy of Sciences, 1784 Sofia, Bulgaria\\
$^{2}$ Irida Observatory, Rozhen, Bulgaria\\
$^{3}$ Department of Theoretical and Applied Physics, University of Shumen, Bulgaria
}

\begin{document}
\date{Accepted \dots Received \dots; in original form \dots}

\maketitle

\label{firstpage}

\begin{abstract}

We obtained and analyzed more than 100 hours of multicolour optical time series of the blazar CTA~102 during its 2012 and 2016 outbursts. The object reached almost 11-th mag at the end of 2016, which is perhaps the brightest blazar state ever observed! During both outbursts, CTA~102 showed significant and rapid variability on intra-night time scales, reaching up to 0.2 mag for 30 min on some occasions. The "\textit{rms}-flux" relation, built for all datasets, shows a large scatter and no apparent saturation on the magnitude scale. The ensemble structure function of the light curves can be fitted well with a straight line of a slope of $\sim$0.4. The time lags between the different optical bands appear to be consistent with zero, taking into account our time resolution. We discuss different variability scenarios and favor the changing Doppler factor of the emitting blobs as the most plausible one to account for the observed intra-night variability.

\end{abstract}


\section{Introduction}

Blazars are jet dominated active galactic nuclei with jet directions presumably closely aligned with the line of sight. Based on the locations of their synchrotron peaks they are broadly classified as low-synchrotron peaked (LSP) and high-synchrotron peaked (HSP) (Abdo et al. 2010). In addition, if accretion disk contribution (thermal peak, broad lines) is present, the blazar is further classified as flat-spectrum radio quasar (FSRQ), almost all of which happen to be LSP. A common feature of practically all types of blazars is their variability over the entire continuum, from radio to X- and gamma-rays. Currently little is known about the exact cause of this variability, but clearly, it is due to processes within the relativistic jet. In the optical, where naturally the blazars are most extensively studied, variations are shown to exist on both long time scales (days--years) and short time scales (hours, so-called intra-night variability). Although not enough statistical evidence is currently available, there are emerging indications that the optical variability is perhaps different for the different blazar types (Hovatta et al. 2014, Gupta et al. 2016). HSPs, on one side, show mostly slow variations in the optical of about a magnitude or two on long time scales (months) and no or very little intra-night variations (e.g. Gaur et al. 2012 [1ES 1959+650, 1ES 2344+514]; Furniss et al. 2015 [Mkn 501]; Aleksic et al. 2015 [Mkn 421]). LSPs and especially FSRQs often show, on the other hand, giant outbursts of up to five magnitudes on long time scales and spectacular intra-night variations, sometimes of $\sim$0.5 magnitudes within several hours (Romero et al. 2000 [AO 0235+164]; Clements et al. 2003 [PKS 0736+017]; Bachev 2015 [S4 0954+65], etc)\footnote{Long term optical light curves of various types of blazars can be found from publicly available archives of Tuorla monitoring program (http://users.utu.fi/kani/1m/), Sankt Petersburg monitoring program (http://vo.astro.spbu.ru/?q=node/10), among others.}. If further statistically confirmed, such a difference in variability may be of significant importance for understanding physical processes and mechanisms leading to blazar emission generation as well as the physical differences between the two major blazar types.
\\
\\
In this work, we explore the optical variations of the FSRQ CTA~102 during its two latest outbursts. Normally, the object stays at a rather low quiescent state of about 17 -- 18 mag; however, in 2012, during an unprecedented back then outburst it reached $\sim$13 mag (Larionov et al. 2016). Even more strikingly, in the late 2016 the object reached $\sim$11 mag (Calcidese et al. 2016). This is an extremely high state not for this blazar only, but also for any blazar. Such bright states (if any) have been reported throughout the years only for very near objects with often-significant host galaxy contribution. CTA~102, on the other hand is a point-like object at $z=1.04$! Had it been at the distance of BL Lacertae (the blazar architype) it would be well visible with a naked eye during its highest states. At the distance of M31, which is perhaps the closest massive galaxy, outside our own, capable of harboring such an object, CTA~102 would almost be the brightest star in our sky! Therefore, such an extraordinary object clearly deserves close inspection and extensive study, especially during its giant outbursts. In this paper, we concentrate on the multicolour intra-night variability of CTA~102 during its 2012 and 2016 maxima. So far, several objects have been multicolour (quasi)-simultaneously monitored on intra-night time scales, among which S4~0954+65 (Papadakis et al. 2004; Bachev et al. 2016), 3C~454.3 (Bachev et al. 2011; Zhai et al. 2011), BL~Lacertae (Zhai \& Wei 2012), S5~0716+714 (Stalin et al. 2009; Wu et al. 2012), etc. To the best of our knowledge, this is the first such study for CTA~102.
\\
\\
The paper is organized as follows: The next section gives details of observations and data reduction. Section 3 is dedicated to the results from our study, Section 4 presents short discussion and next are the conclusions.

\section{Observations and reductions}

CTA~102 was monitored for more than 100 hours on 42 different occasions, mostly during the maximum state, i.e. Sept -- Oct 2012 and June 2016 -- Jan 2017. Repeated exposures in different wavebands were taken during each observational run. Exposure times varied between 60 and 120 sec, chosen to keep photometric errors typically below 0.02 mag. The instruments in use were the 30cm Ritchey Chretien Astrograph of Irida Observatory, Rozhen, Bulgaria, equipped with ATIK 4000M CCD and Sloan g', r', i' filters; the 50/70cm Schmidt camera of Rozhen Observatory, Bulgaria and the 60cm Cassegrain telescope of Belogradchik observatory, Bulgaria. The latter two instruments were equipped with FLI PL 16803 and PL 9000 CCD, respectively, and standard BVRI filter sets.
\\
\\
Standard reduction and aperture photometry procedures were applied in order to extract the instrumental magnitudes. Nearby standard stars were used for the magnitude calibration. Table 1 presents the log of the observations. The first three columns show the evening date (dd.mm.yy format), the instrument and the filters used for the monitoring. The next one gives the duration of the observation in hours. The fifth column shows the estimated average R-band magnitude of CTA~102 during the observation. We used nights when the object was monitored simultaneously with more than one telescope to rescale the Sloan r'-band to the Johnson-Cousins R-band magnitude. Similar rescaling approach was used for the few nights when the objects was observed only in g' or V. We estimate the typical uncertainty of this approach to be of less than 0.05 mag, compared to almost 5 mag overall change (see the next section). The last column gives the root-mean-square (\textit{rms})\footnote{Defined as $\sigma_{\rm rms} = \sqrt{\frac{1}{N-1}\sum_{i=1}^{N} (m_{i}-<m>)^{2} - <\sigma_{\rm phot}>^{2}}$, where $m_{i}$ are the individual magnitude measurements, $<m>$ is the average magnitude, $N$ is the number of data points, and $<\sigma_{\rm phot}>$ is the average photometric uncertainty. $\sigma_{\rm rms}$ is taken to be zero if the expression under the square root happens to be negative.} of the variability amplitude in magnitudes of the R-band light curve, after correcting for the photometric errors. As no significant difference was found in the \textit{rms}'s in different bands, again, proxies were used for the nights with no R-band observations available.
\\
\\
In addition, spectral observations of CTA~102 were carried out with the 2m RCC telescope of Rozhen National Observatory during 3 nights: Dec 22 and 23, 2016, and Jan 02, 2017. Spectral data were reduced using the standard IRAF routines.

\begin{table}
 \centering
  \caption{Observations}
  \begin{tabular}{@{}llcccc@{}}
  \hline
\hline

Date &	Telescope	&	Bands	&	Dur. [h]	&	$<R>$ [mag]	&	rms [mag]	 \\

\hline
21.09.12	&	Bel 60	&	VRI	&	3.3	&	14.91	&	0.105	 \\
22.09.12	&	Bel 60	&	VRI	&	4.0	&	13.82	&	0.114	 \\
05.10.12	&	Bel 60	&	R	&	2.5	&	15.14	&	0.014	 \\
06.10.12	&	Bel 60	&	R	&	2.5	&	15.51	&	0.039	 \\
08.10.12	&	Roz 50/70	&	R	&	1.8	&	15.57	&	0.000	 \\
11.10.12	&	Roz 50/70	&	R	&	1.5	&	15.43	&	0.000	 \\
13.10.12	&	Roz 50/70	&	R	&	0.6	&	15.11	&	0.000	 \\
18.10.12	&	Bel 60	&	VRI	&	3.5	&	15.46	&	0.015	 \\
21.10.12	&	Bel 60	&	VRI	&	2.2	&	15.35	&	0.032	 \\
25.10.12	&	Roz 50/70	&	R	&	0.8	&	15.48	&	0.000	 \\
26.10.12	&	Roz 50/70	&	R	&	0.4	&	15.46	&	0.000	 \\
09.06.16	&	Bel 60	&	VRI	&	0.7	&	15.03	&	0.017	 \\
17.11.16	&	IRIDA 30	&	g'i'	&	2.1	&	13.00	&	0.107	 \\
18.11.16	&	IRIDA 30	&	g'r'i'	&	4.5	&	13.63	&	0.041	 \\
19.11.16	&	IRIDA 30	&	g'r'i'	&	4.2	&	13.69	&	0.034	 \\
20.11.16	&	IRIDA 30	&	g'r'i'	&	4.6	&	13.39	&	0.038	 \\
21.11.16	&	IRIDA 30	&	g'r'i'	&	4.5	&	13.35	&	0.040	 \\
22.11.16	&	IRIDA 30	&	g'r'i'	&	4.7	&	12.56	&	0.082	 \\
23.11.16	&	IRIDA 30	&	g'r'i'	&	4.3	&	12.95	&	0.013	 \\
24.11.16	&	IRIDA 30	&	g'r'i'	&	4.4	&	13.05	&	0.094	 \\
25.11.16	&	IRIDA 30	&	g'r'i'	&	4.6	&	13.32	&	0.027	 \\
26.11.16	&	IRIDA 30	&	g'r'i'	&	0.6	&	13.56	&	0.021	 \\
30.11.16	&	Bel 60	&	BVRI	&	4.9	&	12.56	&	0.080	 \\
03.12.16	&	IRIDA 30	&	g'r'i'	&	0.4	&	12.35	&	0.007	 \\
03.12.16	&	Bel 60	&	BVRI	&	5.3	&	12.37	&	0.090	 \\
06.12.16	&	IRIDA 30	&	g'r'i'	&	2.8	&	12.45	&	0.010	 \\
08.12.16	&	IRIDA 30	&	g'r'i'	&	3.7	&	12.30	&	0.064	 \\
10.12.16	&	IRIDA 30	&	g'r'i'	&	3.5	&	12.90	&	0.037	 \\
11.12.16	&	IRIDA 30	&	g'r'i'	&	2.9	&	12.66	&	0.039	 \\
14.12.16	&	IRIDA 30	&	g'r'i'	&	1.6	&	11.92	&	0.066	 \\
17.12.16	&	IRIDA 30	&	g'r'i'	&	3.0	&	12.70	&	0.044	 \\
18.12.16	&	IRIDA 30	&	g'r'i'	&	0.8	&	12.78	&	0.019	 \\
19.12.16	&	IRIDA 30	&	g'r'i'	&	2.8	&	11.75	&	0.054	 \\
21.12.16	&	IRIDA 30	&	g'r'i'	&	1.6	&	12.10	&	0.037	 \\
21.12.16	&	Bel 60	&	VRI	&	3.4	&	12.07	&	0.124	 \\
22.12.16	&	IRIDA 30	&	g'r'i'	&	1.0	&	11.57	&	0.020	 \\
22.12.16	&	Roz 50/70	&	VI	&	0.8	&	11.46	&	0.025	 \\
23.12.16	&	IRIDA 30	&	g'r'i'	&	0.6	&	11.44	&	0.021	 \\
23.12.16	&	Bel 60	&	V	&	2.2	&	11.49	&	0.049	 \\
23.12.16	&	Roz 50/70	&	BVRI	&	2.5	&	11.43	&	0.032	 \\
24.12.16	&	Bel 60	&	VRI	&	3.4	&	11.96	&	0.101	 \\
02.01.17	&	IRIDA 30	&	g'r'i'	&	1.8	&	12.79	&	0.018	 \\

\hline
\hline
\end{tabular}
\end{table}

\section{Results}

\subsection{Light curves}

Some of the most spectacular variability events are presented in Figures 1 -- 5. Overall changes of up to 0.4 -- 0.5 mag can be seen in the course of a few hours. On some occasions, changes of about 0.2 mag can be traced for as short as 30 min. The character of the intra-night variability varies; from slow trends and wobbles to very rapid and erratic ups and downs (e.g. the night of 22.11.2016, Figure 2).

\begin{figure*}
 \includegraphics[width=120mm]{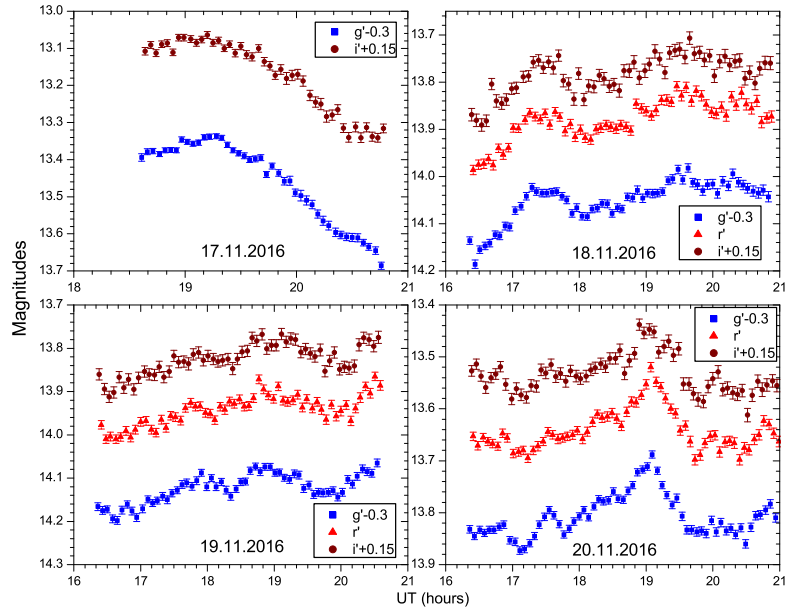}
 \caption{Selected CTA~102 multicolour light curves (see Table 1 for details). The time is UT and the date is shown in every panel. Offsets are applied for presentation purposes. Data from IRIDA Observatory.}
 \label{f1a}
\end{figure*}

\begin{figure*}
 \includegraphics[width=120mm]{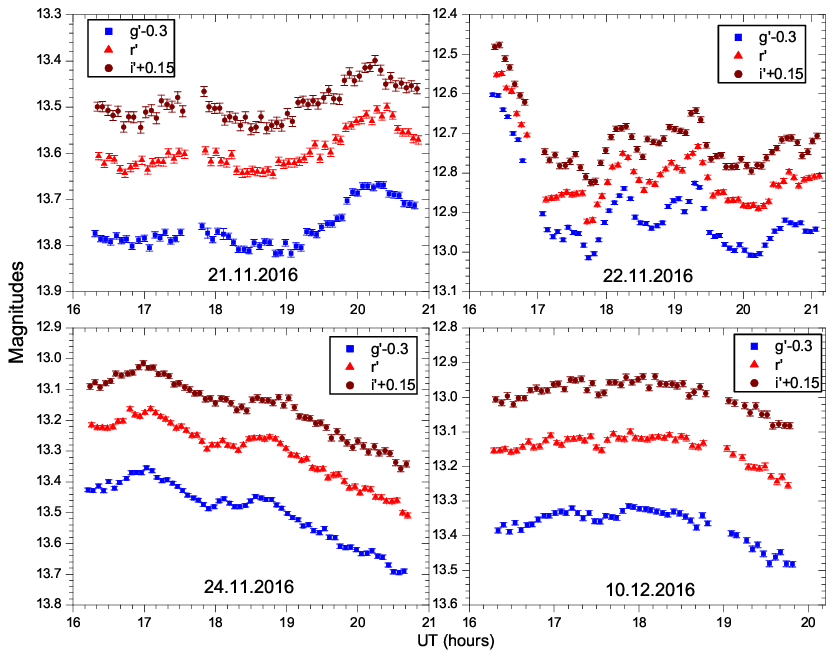}
 \caption{See Figure 1.}
 \label{f1b}
\end{figure*}

\begin{figure*}
 \includegraphics[width=120mm]{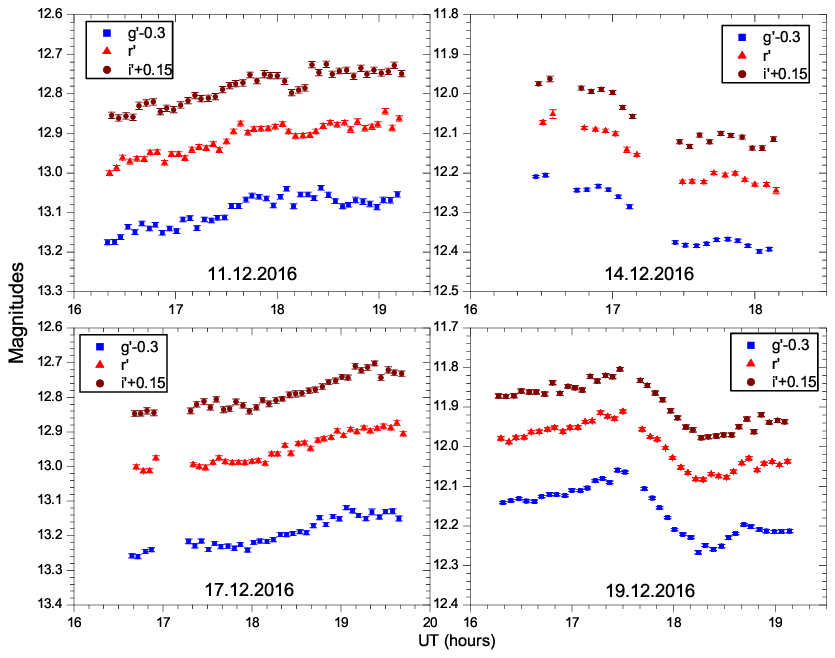}
 \caption{See Figure 1.}
 \label{f1c}
\end{figure*}

\begin{figure*}
 \includegraphics[width=120mm]{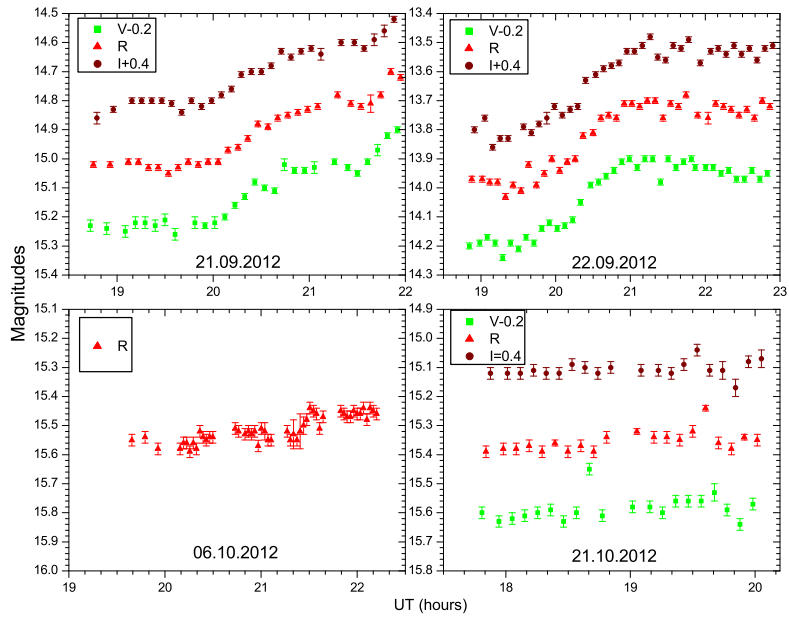}
 \caption{The same as Figure 1. Data from Belogradchik Observatory.}
 \label{f1d}
\end{figure*}

\begin{figure*}
 \includegraphics[width=120mm]{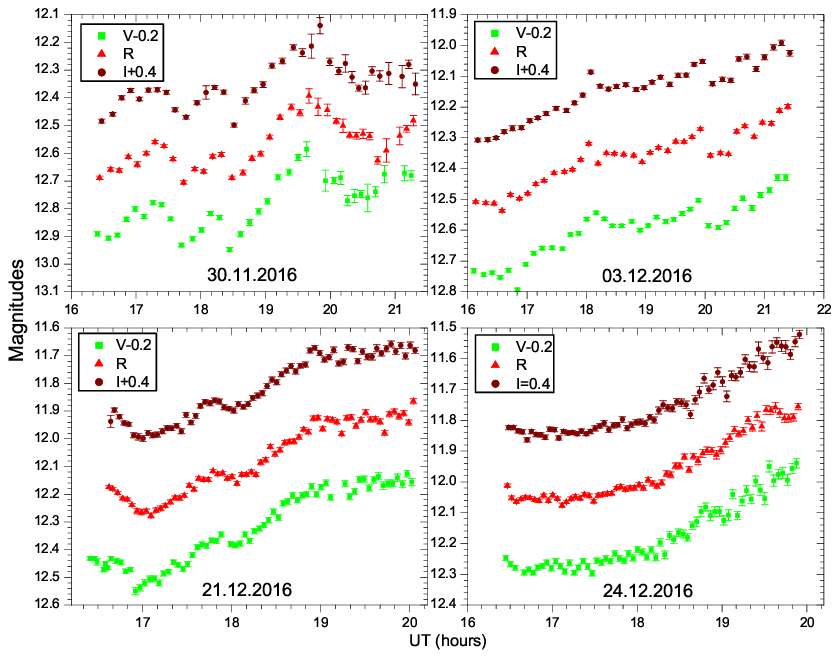}
 \caption{See Figure 4.}
 \label{f1e}
\end{figure*}


\subsection{The "\textit{rms}-flux" relation}

One way to characterize the variability is through its root-mean-square (\textit{rms}, Table 1) of the variability amplitude. 
It has been noticed that for many accreting objects the \textit{rms} is flux dependent (Uttley et al. 2005 [AGNs]; Van de Sande et al. 2015 [CVs]; (Zamanov et al. 2015) [symbiotic stars]; etc.). Figure 6 presents \textit{rms} vs. $<F_{\rm R}>$ for each dataset as blue squares. 
In order to account for the time duration of the datasets, taking into account that the Structure Function (see Sect. 3.4) appears to rise linearly, we calculated in addition the \textit{rms} values for each $\sim$1.5 hour segment of the longer datasets and discarded the shorter ones (Figure 6, red circles).
Both quantities show large scatter in the "\textit{rms}--$<F_{\rm R}>$" diagram, very low correlation, but also -- no clear signs of overall diminishing \textit{rms} trend with the average flux (see the Discussion section).

\begin{figure}
 \includegraphics[width=90mm]{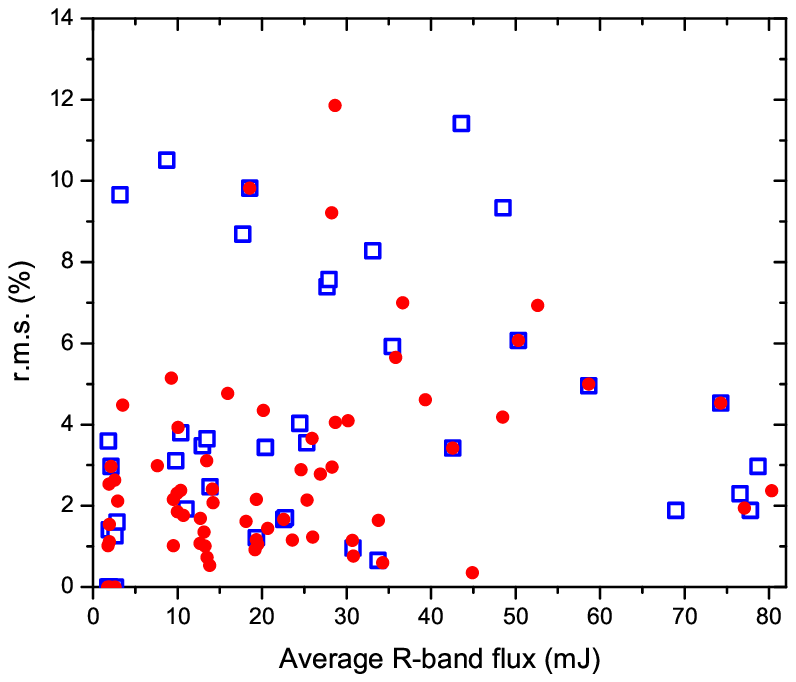}
 \caption{The "\textit{rms}--flux" relation. The blue squares are the actual \textit{rms} values for each dataset.  The red circles are the \textit{rms} for the same datasets binned into $\sim$1.5 hour segments (see the text). 
Note the large scatter and the apparent lack of trends for both quantities.}
 \label{f2}
\end{figure}

\subsection{Time delays}

Monitoring CTA~102 quasi-simultaneously in different colours allows searching for time delays between bands by cross-correlating one with respect to the other as a function of the time displacement between them -- $\tau$. The displacement where the cross-correlation reaches maximum is the time lag between the curves. Our sampling was rather regular, yet some gaps were present. In order to account for the gaps and irregularities, we applied the interpolation cross-correlation method (Gaskell \& Sparke 1986) on the light curves. Clearly, such an approach for searching time delays will work best if the light curves possess some "structure", e.g. multiple maxima and minima present would be the ideal case. We chose nine datasets that appeared suitable for the purpose. Figure 7 presents the results. As seen, most of the datasets are consistent with zero lag. On a few occasions there appear to be small positive or negative lags (here "positive" means the second, longer wavelength band is lagging) of about 2 -- 3 min; however one is to bear in mind that our cadence is about 4 min or more, i.e. we simply do not have enough time resolution to claim any lags, not being consistent with zero.

\begin{figure*}
 \includegraphics[width=140mm]{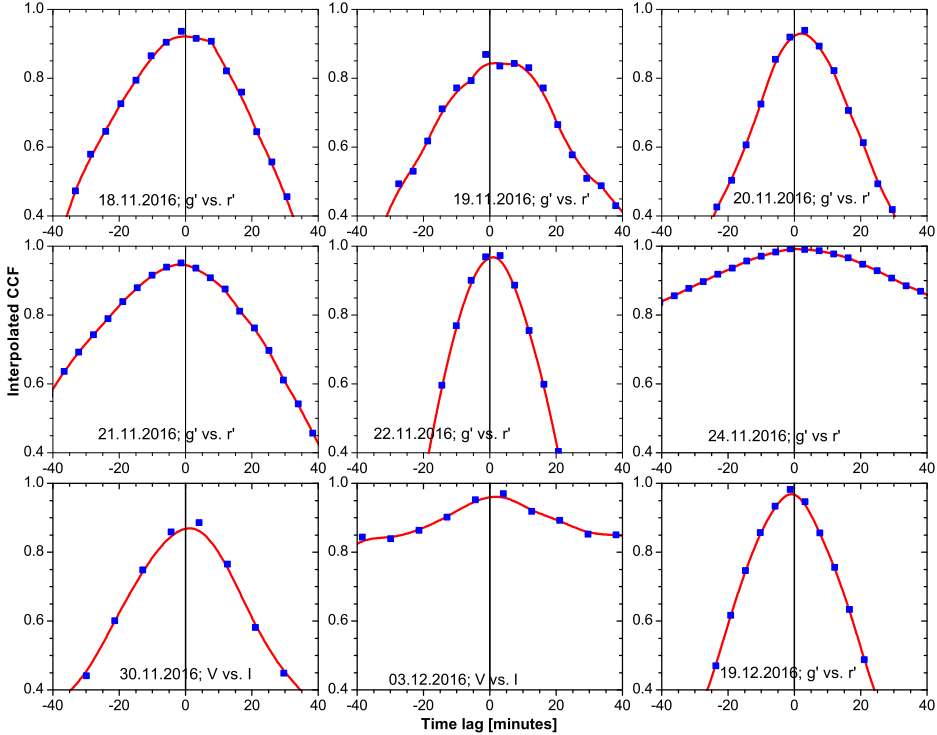}
 \caption{Interpolated cross-correlation functions for selected datasets. Blue dots are the actual correlation values and the red continuous curves are splines to guide the eye. The wavebands used are indicated and are selected based on the minimum photometric uncertainties. A positive lag means the first band is leading.  No significant time lags are found considering the time resolution of our datasets of about 4 min or more. Note the very high correlation (almost one at zero lag) between the bands on some occasions.}
 \label{f3}
\end{figure*}

\subsection{Structure function and time asymmetry}

Figure 8 (upper panel) shows the ensemble first order structure function (SF; Simonetti et al. 1985) for all 42 datasets. 
For a magnitude time series $m(t_{\rm i})$, the SF is defined as $SF(\tau)=\frac{1}{N(\tau)}\sum_{i<j} [m(t_{\rm i})- m(t_{\rm j})]^{2}$, where summation is made over all pairs in which $\tau - \Delta\tau/2 < t_{\rm j} - t_{\rm i} \leq \tau + \Delta\tau/2$, $N(\tau)$ denotes a number of such pairs, and $\Delta\tau$ is the width of the time bin.
The SF is useful to reveal how much the magnitude changes as a function of the time difference between two observations 
(i. e. it is a curve of growth of variability with time lag)
and is normally applied to unevenly sampled data. Saturation (or a slope change) of the SF might indicate the presence of a characteristic time of the system. The ensemble SF of CTA~102 can be successfully fitted with a straight line ($\log\rm{SF} \propto 0.4log\tau$) up to about 3 hours on the time scale (observer's frame), meaning that no characteristic time, shorter than that, could be positively identified. The lower panel of Figure 8 shows the ensemble time symmetry of the light curves. These are calculated as normalized difference between the SF's, built separately for the positive and negative changes of the light curves (Kawaguchi \& Mineshige 1998). The mean and the error of the mean for each time bin are shown. There is some indication, though not too much statistically significant, for the presence of a positive asymmetry at $\tau\ge1$ h. If to be believed, this would mean a small prevalence of rapid rises and gradual decays against the opposite.

\begin{figure}
 \includegraphics[width=90mm]{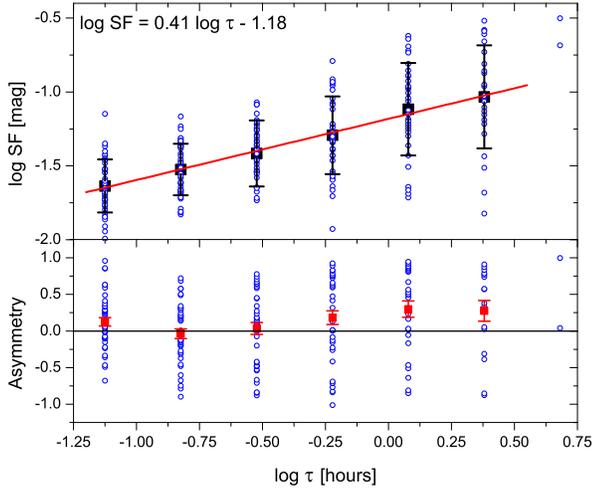}
 \caption{Upper panel: Ensemble structure function, built for all datasets (blue circles). The black error bars show average and the standard deviation of the sample for each time bin. The red continuous line is the best linear fit. Lower panel: Ensemble time symmetry of the light curves (blue circle). The average asymmetry (see the text) and the error of the mean are given as red error bars for each time bin. "Positive" asymmetry here means prevalence of rapid rises and gradual decays against the opposite. Some indications for a positive asymmetry are seen for $\tau\ge1$ hour but in general the light curves appear to be fairly time symmetric.}
 \label{f4}
\end{figure}

\subsection{Optical spectrum}
Figure 9 shows an optical spectrum of CTA~102, taken on 23.12.2016, when the object was around 11.5 mag (R-band).  Interestingly, no MgII$\lambda$2798 line can be seen, being either weak or most probably -- concealed in the noise. MgII is clearly seen on previous spectra (Larionov et al. 2016), where, however, the object was much fainter. A quick check with the Steward Observatory public blazar archive\footnote{http://james.as.arizona.edu/$\sim$psmith/Fermi/} shows that their spectra taken at about that time also lack the emission line.
 
\begin{figure}
 \includegraphics[width=90mm]{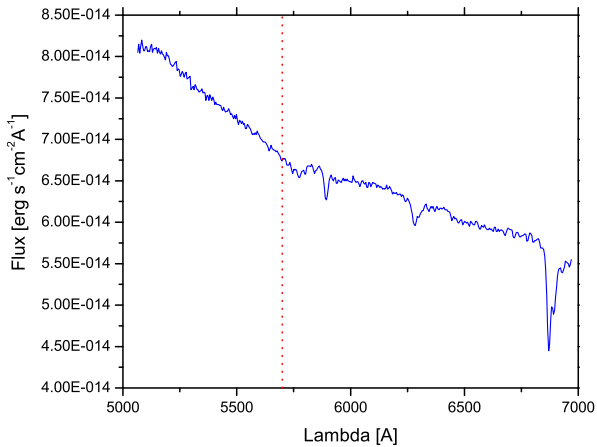}
 \caption{Optical spectrum of CTA~102 taken on 23.12.2016 with the 2m RCC telescope of Rozhen Observatory. The position of the MgII$\lambda$2798 emission line is indicated with a red dashed line. The line is clearly absent, perhaps concealed in the continuum noise during this very high state.}
 \label{f5}
\end{figure}

\section{Discussion}

As the exact drivers of blazar variability, especially on intra-night time scales, are still under debates, any detail study of such should be of importance to help solving the problem. This is especially true when the object of interest shows indeed strong and rapid variations, as it is for the CTA~102 case. Light curves alone, are perhaps not enough to understand the entire physics of the blazar jets and fully solve the variability problem but they can help at least to restrain the models. What can we learn from our analysis in that respect?
\\
\\
There are several mechanisms to cause variations of blazar emission. The most promising of them include microlensing, fast evolution of the electron energy-density distribution and changes of the Doppler factor (so called geometrical reasons). Although the microlensing from intervening bodies along the line of sight cannot be fully excluded, it is unlikely to play significant role on intra-night time scales, due to the very fast changes observed. Furthermore, it is unclear why microlensing should play significant role only for LSP objects (including FSRQs), as the presence of intervening bodies should not be related to the type of the object. Fast acceleration and subsequent energy loss of the emitting particles could in principle produce variations, perhaps even on intra-night time scales. Marscher (2014), for instance, demonstrated the possibility for such a rapid variability using his Turbulent Extreme Multi-Zone (TEMZ) model, where a large number of turbulent cells pass through a standing conical shock wave. In that case, however, one should expect clear time delay between the bands, as the higher energy particles evolve faster. Such a delay can roughly be expressed (following B{\"o}ttcher, 2007) as $\tau\simeq5B^{-3/2}\delta_{10}^{-1/2}(1+u_{ph}/u_{B})^{-1}(\sqrt{\lambda_{2, 5000}}-\sqrt{\lambda_{1, 5000}})$ [hours], 
where $\lambda_{1-2, 5000}$ are the two wavelengths in units of 5000$\AA$, $B$ is the magnetic field, $\delta_{10}=\delta/10$, $u_{ph}$ and $u_{B}$ are the photon and magnetic field energy densities respectively. Thus, the time delay expected can be as long as a few hours.  We should be able to detect such delays even with our moderate time resolution, but we did not. This result is not really surprising as the most of the other studies fail to detect time lags between the optical bands or, if found, such lags are either rather uncertain or observed only in particular epochs. For instance, for S5~0716+714, which is the most extensively studied in terms of multicolour intranight variability object, the results show either no lags or such of low statistical significance due to large errors (e.g. Villata et al. 2000; Qian et al. 2000; Stalin et al. 2006; Zhang et al. 2010; Hao et al. 2010). Wu et al. (2012), however, claimed a clear delay of R-band behind V-band of about 30 min but observed only in one night during their campaign. Similar results (lags generally consistent with zero considering the errors) were reported for BL Lacertae (e.g. Papadakis et al. 2003; Stalin et al. 2006; Zhai \& Wei 2012) and S4~0954+65 (Papadakis et al. 2004; Bachev et al. 2016). A possible lag was found for one night of observations of 3C~454.3 during two different campaigns (Zhai et al. 2011; Bachev et al. 2011) and for all the other nights the lag was practically consistent with zero. Romero et al. (2000) found, on the other hand, a small ($\sim$4 min) lag of R-band variations relative to the V-band ones for PKS~0537-441. Clearly, if time lags between the optical bands exist at all, the results do not appear to be systematic in any way. The problem with many of the studies mentioned above is that they often rely on small-scale variability when deriving the lags between the light curves, where, due to the photometric noise often the errors in the CCFs lead to uncertainties of the time lags as large as the lags themselves. In other words, a light curve that shows barely any variability (considering the photometric errors) or shows just trends is hardly suitable for this purpose. That is why we considered only about 20 percent of our datasets suitable for this analysis. In addition, an energy injection and the consequent cooling would in principle lead to some time asymmetry of the light curves as there is no a priori reason for the injection and cooling times to be exactly equal.
\\
\\
To our opinion, the fast intra-night variations observed can be best explained by invoking the geometrical reasons. These require disordered (curved) magnetic field along which the emitting blobs (or more generally -- active regions) are travelling. At some point the direction of the blob (or several of them) happens to coincide closely to the line of sight, leading to a significant enhancement of the emitted light. Indeed, taken into account that the Doppler factor $\delta= \frac{1}{\Gamma(1-\beta \cos\theta)}$ and observed-to-emitted flux ratio is $\propto\delta^{3+\alpha}$, where $\alpha$ is the spectral index, one can see that only a small change of the direction of the blob with respect to the observer, $\theta$ (say from 4 to 0 degrees), can produce almost a hundred times emitted flux change\footnote{Provided the bulk Lorentz factor $\Gamma\simeq20$ as often assumed for blazars like CTA 102 and $\alpha\simeq1$}.
\\
\\
Doppler factor changes should produce symmetric in time light curves, which seems to be confirmed by our findings, as no significant asymmetry was really found. There are two ways to explain the observed "\textit{rms}--flux" relation. 
Let first assume that the overall brightness level is controlled by the number ($N$) of the active regions, being currently present. Presumably, they are launched at the jet base and travel down the (curved) jet. At some point their Doppler factor would reach maximum, and the corresponding active region will "switch-on", thus starting to contribute significantly toward the overall brightness. For a 5+ magnitude outburst, such as the 2016 one, however, $N$ must be significant (100+) at least during that outburst, and therefore the Large Number Theorem (LNT) should become applicable. This means that expected flux $rms\propto1/\sqrt{N}$ and should decrease for high brightness states, something we find no clear signs of (Figure 6). Similar results were also reported by Bachev et al. (2016) for S4~0954+65. However, there is a caveat to be mentioned. LNT applies to so-called additive processes, for which the number of active regions launched within a fixed time period is Poisson (Gaussian) distributed. There are, on the other hand, multiplicative processes, such as the avalanche-type processes (Aschwanden 2014), which are not necessarily Gaussian distributed and the LNT is not necessarily applicable there. They in principle will lead to a different "\textit{rms}--flux" relation, perhaps similar to the one we observed. Such a process is described by Kawaguchi et al. (1998), indeed not for jets but for accretion disk instabilities. In their scenario, if the number of the active regions happens to exceed some certain value, new active regions are generated, thus leading to the formation of an avalanche-type process. Interestingly enough, our relatively shallow SF slope well agrees with their multiplicative (otherwise unrelated to our case) disk-instability model, in contrary to the models of random explosive events, where the slope appears to be much steeper. In other words, one way to account for the observed variability is to assume multiple active regions, constantly switching on and off, due to the changing Doppler factor (ultimately leading to the intra-night variability), while their gradually changing average number, due perhaps to accretion rate changes, is responsible for the long-term variations, such like the giant the outbursts.
\\
\\
On the other hand, there is another scenario that does not require significantly changing average $N$ to account for the low and high states, but instead, relies on the change of the direction of the entire jet cone. When the jet cone is closely aligned with the line of sight there would be an outburst and vice versa. In such a case, the intra-night variability properties, including the "\textit{rms}--flux" relation, are virtually the same for all states; the overall brightness is just hugely amplified when a close aligning occurs. The entire jet may change its direction due perhaps to a precession of the black hole, provided the black hole and accretion disk spin momenta are misaligned. Although Nixon \& King (2013) recently explored such a possibility and argued that the black hole will "hardly move", there is still a possibility for the accretion disk to precess due to the Lense-Thirring effect. In that respect, note that there are still debates what generates the jet, black hole spin (Blandford-Znajek effect; Blandford \& Znajek 1977) or accretion disk itself (Blandford-Payne effect; Blandford \& Payne 1982). Whichever case, it is clear that the FSRQ, powered by a thin (presumably misaligned) accretion disk are more likely to produce precessing or at least wobbling jets than the other types of blazars, which are likely powered by geometrically thick flow (ADAF?), where naturally the matter should be more uniformly distributed around the black hole equator. This may be a plausible explanation why the FSRQs seem more likely to produce giant outbursts, while the other blazars -- do not.
\\
\\
The absence (or the extreme weakness) of MgII$\lambda$2798 line at high states may also imply that the gamma-ray production site is located far from the broad line region, since otherwise some response of the line during the outburst should be expected. Such a response was reported for e.g. 3C 454.3 (Leon-Tavares et al. 2013), but for CTA~102 the MgII line seems to stay constant in flux all the time (Larionov et al. 2016) and appear to vanish in the continuum noise only during very high states. Alternatively, MgII can really fade in flux, if suppose that at some moment the entire jet cone is directed towards the observer. Then, all the photons from the broad line region in our direction will have to cross the jet cone at some point and a large portion of them would be up-scattered and shifted towards the higher energies (thus being lost from the optical region).

\section{Conclusions}
We present the results of over 100 hours of multicolour optical monitoring of the FSRQ CTA~102 during its 2012 and 2016 unprecedented outbursts. The object showed rapid and significant in amplitude intra-night variations for the most of the observational epochs, reaching occasionally up to 0.2 mag variations for about 30 min. We find no detectable time lags between the optical bands. The ensemble structure function shows no signs of saturation or slope change for a time span shorter than about 3 hours. Light curves appear symmetric in time or no statistically significant evidence for the opposite is found.
\\
\\
We favor geometrical factors (changing Doppler factor of blobs, moving along the jet) as a plausible explanation of the observed variability picture. If the total number of blobs controls the average brightness level, then they should be launched via multiplicative (e.g. avalanche-type) processes at the jet base, to account for the observed "\textit{rms}--flux" relation. Alternatively, the brightness level may be controlled by a change of the direction of the entire jet cone, due perhaps to a precession. The latter possibility may explain why FSRQs, powered particularly by a thin and possibly -- misaligned accretion disk, appear to produce more frequently giant outbursts.

\section*{Acknowledgments}
This research was partially supported by the Bulgarian National Science
Fund of the Ministry of Education and Science under grants DN 08-1/2016
and DM08-2/2016 and by funds of the project RD-08-81 of Shumen University. We thank the anonymous referee for carefully reading the text and the useful suggestions that helped to improve the paper.

\end{document}